\documentclass[aps,prapplied,twocolumn,superscriptaddress]{revtex4-2}
\usepackage{amsmath}
\usepackage{color}
\usepackage[dvipsnames]{xcolor}
\usepackage{graphicx}
\usepackage{dcolumn}
\usepackage{bm}
\usepackage{braket}
\usepackage{xcolor}
\usepackage[normalem]{ulem} 
\usepackage[detect-weight=true,separate-uncertainty=true]{siunitx} 

\usepackage{cleveref}
\crefname{equation}{Eq.}{Eqs.}
\Crefname{equation}{Equation}{Equations}
\crefname{figure}{Fig.}{Figs.}
\Crefname{figure}{Figure}{Figures}
\crefname{section}{Sec.}{Sects.}
\Crefname{section}{Section}{Sections}
\crefname{table}{Table}{Tables}
\crefname{appendix}{Appendix}{Apps.}
\Crefname{appendix}{Appendix}{Apps.}

\def \hrho{\hat{\rho}}
\def \ha{\hat{a}}

\def \hf{\hat{f}}
\def \hn_t{\hat{n}_t}
\def \hphi_t{\hat{\varphi}_t}
\def \hH{\hat{H}}

\begin{document}


\title{High-fidelity gates in a transmon using bath engineering for passive leakage reset}


\author{Ted Thorbeck}
\email{ted.thorbeck@ibm.com}
\affiliation{IBM Quantum, IBM T.J. Watson Research Center, Yorktown Heights, NY 10598, USA}
\author{Alexander McDonald}
\affiliation{Institut Quantique and D\'epartement de Physique, Universit\'e de Sherbrooke, Sherbrooke J1K 2R1 QC, Canada}
\author{O. Lanes}
\affiliation{IBM Quantum, IBM T.J. Watson Research Center, Yorktown Heights, NY 10598, USA}
\author{John Blair}
\affiliation{IBM Quantum, IBM T.J. Watson Research Center, Yorktown Heights, NY 10598, USA}
\author{George Keefe}
\affiliation{IBM Quantum, IBM T.J. Watson Research Center, Yorktown Heights, NY 10598, USA}
\author{Adam A. Stabile}
\affiliation{IBM Quantum, IBM T.J. Watson Research Center, Yorktown Heights, NY 10598, USA}
\author{Baptiste Royer}
\affiliation{Institut Quantique and D\'epartement de Physique, Universit\'e de Sherbrooke, Sherbrooke J1K 2R1 QC, Canada}
\author{Luke C. G. Govia}
\affiliation{IBM Quantum, IBM Almaden Research Center, San Jose, California 95120, USA}
\author{Alexandre Blais}
\affiliation{Institut Quantique and D\'epartement de Physique, Universit\'e de Sherbrooke, Sherbrooke J1K 2R1 QC, Canada}
\affiliation{Canadian Institute for Advanced Research, Toronto, M5G 1M1 Ontario, Canada}

\date{\today}

\begin{abstract}
Leakage, the occupation of any state not used in the computation, is one of the of the most devastating errors in quantum error correction. Transmons, the most common superconducting qubits, are weakly anharmonic multilevel systems, and are thus prone to this type of error. Here we demonstrate a device which reduces the lifetimes of the leakage states in the transmon by three orders of magnitude, while protecting the qubit lifetime and the single-qubit gate fidelties.  To do this we attach a qubit through an on-chip seventh-order Chebyshev filter to a cold resistor. The filter is engineered such that the leakage transitions are in its passband, while the qubit transition is in its stopband.
Dissipation through the filter reduces the lifetime of the transmon's $f$ state, the lowest energy leakage state, by three orders of magnitude to 33~ns, while simultaneously keeping the qubit lifetime to greater than 100~$\mu$s.  Even though the $f$ state is transiently populated during a single qubit gate, no negative effect of the filter is detected with errors per gate approaching \num{1e-4}. Modelling the filter as coupled linear harmonic oscillators, our theoretical analysis of the device corroborate our experimental findings. This leakage reduction unit turns leakage errors into errors within the qubit subspace that are  correctable with traditional quantum error correction. We demonstrate the operation of the filter as leakage reduction unit in a mock-up of a single-qubit quantum error correcting cycle, showing that the filter increases the seepage rate back to the qubit subspace. 
\end{abstract}

\pacs{Valid PACS appear here}
\maketitle

\section{Introduction}

Leakage to states outside the computation subspace is common in weakly anharmonic qubits like transmons, the most commonly used superconducting qubit \cite{koch2007transmon}.
Even basic operations such as one- and two-qubit gates and measurement are known to cause leakage \cite{wood2018quantification, sank2016measurement}.  This type of error is one of the most serious threats to quantum computation because quantum error correction (QEC) algorithms do not correct for leakage errors. Minimizing leakage often comes at the cost of slowing down the operation, thus increasing decoherence-related errors. Waiting for leakage states to relax back to the qubit subspace, the most common method to remove leakage, has become an untenable solution as qubit lifetimes have increased, which has also increased the lifetimes of the leakage states. Furthermore, postselection, in which any data that shows signs of leakage is discarded, has begun to fail as a mitigation approach because the  odds of a leakage-free computation have shrunk as the number of qubits and operations has increased; postselection now often requires discarding most of the data in a large computation \cite{sundaresan2023demonstrating, varbanov2020leakage}. Therefore, while recent increases in coherence times and the number of qubits used in superconducting quantum computers are enabling more complex quantum computations,  these same increases are also making leakage a more dangerous threat to superconducting quantum computation \cite{sundaresan2023demonstrating, krinner2022realizing, google2023suppressing}. To build a quantum computer, either leakage events must be eliminated or another approach is needed to cope with this type of error. 

A leakage reduction unit (LRU) takes the second approach,  converting leakage errors into errors that can be corrected by QEC algorithms by quickly returning leakage back to the qubit subspace \cite{aliferis2005fault, suchara2015leakage}.  In superconducting qubits, most LRUs require an active step during the QEC cycle during which leaked transmon states are returned to the qubit subspace.
For example, if a measurement detects the occupation of the $f$ state, the first state outside the qubit subspace, then a conditional $\pi$-pulse at $\omega_{ef}$ can be applied to return the qubit to the $e$ state \cite{battistel2021hardware}. However, in a QEC algorithm only the ancilla qubits are routinely measured, so this approach cannot be used on the data qubits.
For data qubits, a microwave-driven Raman transition can swap population from the $f$ state of the qubit into the readout resonator, where it quickly decays \cite{battistel2021hardware, marques2023all, egger2018pulsed, magnard2018fast, lacroix2023fast, zeytinouglu2015microwave}. While both of these methods work for the $f$ state, and sometimes the $h$-state \cite{lacroix2023fast, battistel2021hardware, marques2023all}, they struggle for leakage to higher levels because of the increasing charge dispersion. Unfortunately leakage to states higher than the $f$ state is common during strong microwave drives used in measurement \cite{sank2016measurement, dumas2024unified, khezri2022measurement, shillito2022dynamics}, microwave-driven two-qubit gates \cite{malekakhlagh2022optimization, malekakhlagh2022mitigating, tripathi2019operation}, and microwave pump tones \cite{verney2019structural, lescanne2019escape}. Reset of such highly excited transmon states has been demonstrated by pulsing a flux-tunable transmon through a lossy mode \cite{mcewen2021removing}.  However, this destroys the qubit state, so it can only be applied to ancilla qubits. This limitation can be overcome by either swapping the quantum information from the data qubits to the ancilla qubits before resetting the data qubits, or by swapping the leakage from the data qubits to the ancilla qubits before resetting the ancilla qubits \cite{miao2023overcoming, ghosh2015leakage, camps2024leakagemobilitysuperconductingqubits}. 
Moreover, each of these methods require time during the quantum algorithm, time during which the other qubits are suffering from decoherence.

The ideal LRU for a superconducting quantum computer would reset any leakage state back to the qubit subspace, not disturb the qubit subspace, not slow down a computation by adding an additional steps the algorithm, work for either flux-tunable or fixed-frequency transmons, and  be completely passive. In this paper we demonstrate an LRU which satisfies all these requirements. It consists of a transmon connected to a resistor through a high-order filter, as shown in \cref{fig:basic_idea}.  The filter is engineered such that the frequencies of the leakage transitions of the transmon are in the passband of the filter, so that the leakage states are short lived. The frequency of the  qubit transition, however, is in the stopband of the filter, protecting the lifetime of the qubit.  We engineered a seventh-order Chebyshev filter so that the roll-off of the filter would be sharper than the small anharmonicity of the transmon. The filter reduces the lifetime of the $f$ state by three orders of magnitude from $T_{1,f}$ = 34~$\mu$s in the stopband of the filter to $T_{1,f}$ = 33~ns in the passband of the filter, which is approaching the duration of a typical single-qubit gate. Nevertheless the lifetime of the qubit's first excited state is over 100~$\mu$s, unaffected by the filter. Even though the $f$ state is transiently occupied during single qubit gates, we  demonstrate high-quality single-qubit gates on the qubit, as measured by randomized benchmarking. We also study the filter from a theoretical perspective, modelling it quantum mechanically and performing full time-dependent simulations. These simulations corroborate our experimental findings that the filter does not significantly degrade our ability to perform quantum computations. Finally, to demonstrate operation as an LRU, we study the effect of the filter on the leakage and seepage rates during a mock-up of a QEC circuit which consists of repeatedly measuring the qubit, often the most leaky operation in a quantum computer~\cite{sundaresan2023demonstrating, sank2016measurement}.

It is not obvious that such a filter can be coupled to a transmon without negative effects on the qubit. In weakly anharmonic qubits like transmons, fast single-qubit gates require large-bandwidth microwave pulses that also drive the $\omega_{ef}$ transition, transiently populating the $f$ state during the gate \cite{steffen2003accurate}. The commonly used derivative removal by adiabatic gate (DRAG) method reduces the population in the $f$ state at the end of the gate, but the $f$ state is still populated during the gate \cite{motzoi2009simple, chow2010optimized, chen2016measuring}. This raises the question: if the $f$ state is weakly transiently-occupied during the gate, how much decoherence on the $ef$-transition can be tolerated before it negatively affects a single-qubit gate?  Therefore we focus on demonstrating high-quality single-qubit coherent operations in a qubit connected to a leakage reset filter.  We also examine how the filter affects single qubit operations on neighboring qubits, but we do not study how the filter affects two-qubit gates. 

High-order filters already have many uses in superconducting quantum circuits, such as in Purcell filters \cite{bronn2015broadband, yan2023broadbandpurcell, park2024characterization, zhang2023superconducting}. 
Similar multi-mode structures can also be referred to as photonic crystals \cite{harrington2019bath}, photonic bandgaps \cite{liu2017quantum}, superconducting metamaterials \cite{mirhosseini2018superconducting}, and waveguides \cite{kim2021quantum}. Such structures have been used to enable long-distance qubit interactions \cite{mckay2015high}, generate qubit interactions with photonic bound states \cite{sundaresan2019interacting}, engineer dissipation \cite{liu2017quantum, harrington2019bath}, engineer dispersion in traveling-wave parametric amplifier \cite{ho2012wideband}, and for quantum simulation \cite{zhang2023superconducting}. Photonic bandgaps has previously been used to suppress of $T_{1,f}$ of a transmon with respect to $T_{1,e}$ \cite{mirhosseini2018superconducting, vadiraj2021engineering, kockum2014designing}, but not with a focus on high-quality qubit operations.  A filter has been previously proposed to reduce leakage in a qubit with a large energy gap to the leakage states \cite{putterman2022stabilizing}, however the engineering of the filter is very different in a small anharmoncity qubit like the transmon.

\section{Engineering the filter}

\begin{figure}[t!]
\includegraphics[]{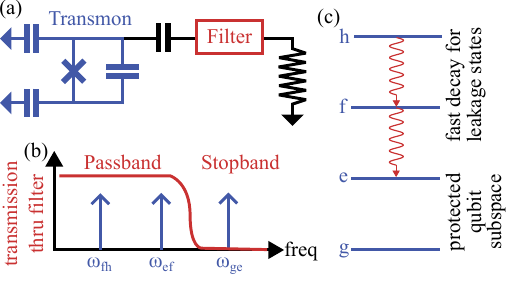}
\caption{ (a) A transmon (blue) is capacitively coupled to resistor (black) thru a filter (red).  (b) The leakage transitions ($\omega_{ef}$, $\omega_{fh}$ ...)  are situated in the passband of the filter, coupling them to the resistor, reducing their lifetimes.  While the qubit transition, $\omega_{ge}$, is in the stopband of the filter, isolating it from the resistor, protecting its lifetime.  Therefore the filter must have a very sharp rolloff so that the transition band of the filter is smaller than the anharmonicity, $\alpha = \omega_{ef} - \omega_{ge}$.  (c) If a transmon has leaked to the $h$-state, the dissipation induced at $\omega_{fh}$ induces relaxation to the $f$ state, which quickly decays to the $e$ state, at which point the cascade stops because $\omega_{ge}$ is protected in the stopband of the filter. Once the transmon has returned to the qubit subspace, the LRU has been successful.
 } 
	\label{fig:basic_idea}
\end{figure}

To act as an LRU the filter must: 1) have a passband broad enough to pass several leakage transitions, 2) a stopband around the qubit frequency, 3) a sharp roll-off to transition from passband to stopband over a transition band smaller than the anharmonicity of the qubit, and 4) minimal ripple in the passband. In this section we design a filter for a typical qubit with $\omega_{ge}/2\pi$ = 4.7~GHz, with $\alpha/2\pi$~=~-325 MHz.  The anharmonicity, which is large for a transmon, makes condition 3 easier to realize and is beneficial for fast cross-resonant gates between qubits \cite{malekakhlagh2020first}.

First, as in Fig.~\ref{fig:basic_idea}(b), the passband should be broad enough to contain multiple leakage transitions ($\omega_{ef}$, $\omega_{fh}$, ...). This is because if the transmon has leaked to the $h$-state, then dissipation from the resistor induced at $\omega_{hf}$ causes the $h$-state to quickly decay to the $f$ state.  It will then quickly relax to the $e$ state, at which point the cascade will stop because $\omega_{ge}$ is protected in the stopband of the filter, see Fig.~\ref{fig:basic_idea}(c). Now the LRU has successfully returned population in the leakage states back to the qubit subspace and a leakage error has successfully been converted to an error within the computational subspace, which can be corrected by QEC. The filter is not designed to provide reset to the ground state from within the qubit subspace, as often used in initialization. However, the filter can in principle be used to initialize the qubit in the $e$ state by driving the two-photon $g-f$ transition, which can then be rotated to the $g$ state with a $\pi$-pulse.

Second, as shown in Fig.~\ref{fig:basic_idea}(b) the qubit transition $\omega_{ge}$ should be in the stopband of the filter, protecting the qubit from dissipation due to the resistor. The stopband should be broad enough to protect the qubit given variations in qubit frequency \cite{zhang2022high}, as well as frequency shifts and broadening during gates and readout \cite{thorbeck2022tls}. Here we choose to implement the filter as a bandpass filter around the leakage transitions. We set the bandwidth of the filter at $\Delta\omega/2\pi$ =  1.8~GHz, to pass roughly the first five leakage transitions, beyond which the charge dispersion of the leakage transitions becomes very large. We engineer the cutoff frequency of the filter to be at 4.5~GHz, in between $\omega_{ge}/2\pi$ and $\omega_{ef}/2\pi$.  Although we focus on the bandpass implementation of the filter other filter typologies would also work: a bandstop filter around the qubit frequency, or a low-pass (high-pass) filter for qubits with negative (positive) anharmonicity.

Third, the filter must have a steep rolloff between the passband and the stopband because $\omega_{ge}$ and $\omega_{fe}$ are detuned by $\left|\alpha\right| \ll \omega_{ge}$.  In general higher-order filters have steeper rolloff, at the expense of a larger physical footprint. Fourth and finally, excessive ripple in the passband of the filter can result in a large spread in the lifetimes of the different leakage transitions.  This is undesirable because the relaxation of leakage is a sequential process, and so is limited by the slowest decay rate. Chebyshev filters offer a good balance of steep rolloff and reasonable ripple in the passband, with parameters that are easy to design and fabricate \cite{pozar2011microwave}. Therefore we based our design on a seventh-order Chebyshev filter with a 0.1 dB ripple factor. 

The final parameters of the LRU, the coupling between the qubit and the filter $g$, and the resistance $R$ determine the lifetime of the leakage transitions in the passband of the filter. We set $R$ = 50~$\Omega$ because we used an off-chip termination as the source of dissipation to simplify manufacturing. Therefore, the desired lifetime of the leakage states sets the value of $g$.  We engineered this LRU to suppress the lifetimes of the leakage states to less than 100~ns.  To act as an LRU the filter likely only needs to reset leakage on a timescale similar to the typical QEC cycle time, which today is about 1~$\mu$s \cite{sundaresan2023demonstrating, google2023suppressing, krinner2022realizing}. This is because recent work in the surface code has shown that the LRU can improve the logical error rate even if the LRU does not perfectly reset leakage during every QEC cycle \cite{varbanov2020leakage}.  We choose the lower lifetime target of 100~ns for this filter to study detrimental effects of the filter on qubit operations.  

\begin{figure}[t!]
	\includegraphics[width=\columnwidth]{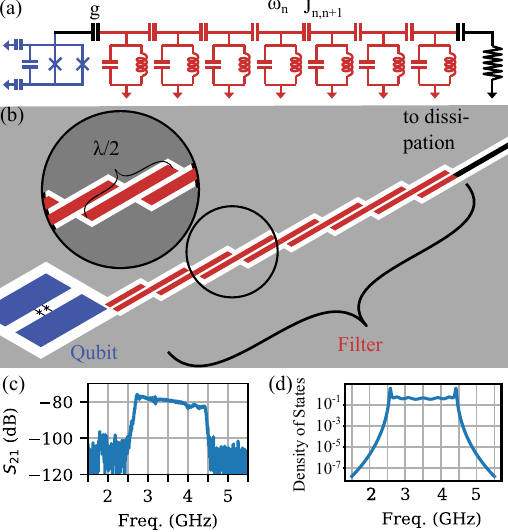}
	\caption{Implementing the filter.  (a) Diagram of the circuit for quantization.  The qubit (blue) is capacitively coupled, with coupling strength $g$, to a filter (red), which is connected to off-chip dissipation (black). To quantize the circuit, the seventh-order filter can be represented with seven coupled lumped-element resonators.  (b) Cartoon of the physical layout of the qubit and the filter, not to scale.  The qubit (blue) is coupled to a coupled-line filter (red) to off-chip dissipation (black).  The coupled mode filter consists of parallel transmission lines.  The modes consists of seven $\lambda$/2 sections, cascaded such that half of each segment overlaps with the prior segment, and the other half overlaps with the subsequent segment, to provide the coupling.  (c) Measured transmission $S_{21}$ through a test structure for the filter, not coupled to any qubits, showing a passband $\sim$~2~GHz broad.  This measurement is not calibrated to remove attenuation and dispersion in drive lines. (d) Density of states (DOS) for the filter  calculated using \cref{eq:DOS}, as a function of frequency.  Note that the regions of high density of states correspond to regions of high transmission through the filter. 
 } 
	\label{fig:filter_phys}
\end{figure}

Once the filter parameters have been determined, the design can be implemented in many ways using standard microwave engineering \cite{pozar2011microwave}.  A canonical form for a $L$th-order bandpass filter consists of $L$ capacitively coupled resonators, as shown in  Fig. \ref{fig:filter_phys}(a). Each resonator individually would act as a single-pole bandpass filter near its resonant frequency, $\omega_i$, and the coupled resonators hybridize forming collective modes and creating a passband. However, the large capacitances and inductances required can be difficult to fabricate. Instead, we fabricated a coupled-line filter implementation, see Fig. \ref{fig:filter_phys}(b). An $L$th-order filter then consists of $L$ half-wavelength, open-circuited, coplanar waveguide (CPW) resonators \cite{pozar2011microwave}. The resonators are cascaded by running the CPWs in parallel to couple each resonator to its neighbors.  The amount of coupling between each resonator, $J_{n,n+1}$, can be engineered by controlling the width of the CPWs, the gaps between the CPWs, and the gap between the CPWs and the ground plane.
Because the purpose of this device is to test the basic concept, we did not focus on reducing the size of the filter.  In addition, lossy materials should be avoided in fabricating the filter; because while the filter protects the qubit from dissipation due to the resistor, the qubit will still partially hybridize with the filter. Therefore internal loss in the filter reduces the lifetime of the qubit.  

\section{Experimental Implementation of the Filter}

\subsection{Measurements of a transmon connected to the filter}

\begin{figure}[t!]
	\includegraphics[width=\columnwidth]{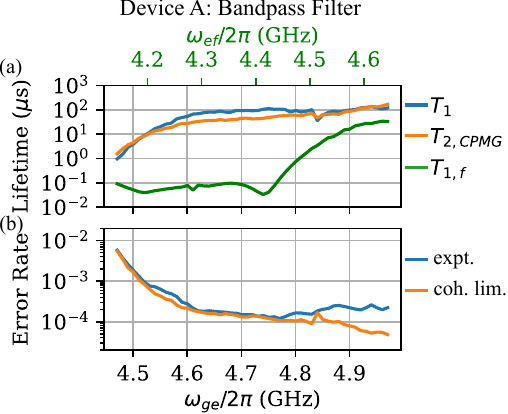}
	\caption{Effect of the filter on the transmon.  (a) Measurements of $T_1$, $T_{2,\mathrm{CPMG}}$, and $T_{1,f}$ as a function of  frequency.  For the $x$-axis we show both $\omega_{ge}$ in black on the bottom of the plot and $\omega_{ef}$ in green on the top of the plot. Although the passband of the filter begins at 4.5~GHz, because of the finite transition band of the filter, $T_{1,e}$ rolls off below 4.6~GHz. However, the $ef$-transition enters the passband of the filter when the qubit frequency is at 4.9~GHz, because of the anharmonicity of the transmon. The filter suppresses $T_{1,f}$ by three orders of magnitude, from 34~$\mu$s to 33~ns, as the qubit frequency changes by only 200~MHz.  Therefore this filter could act as an LRU for qubits frequencies between 4.6~GHz and 4.8~GHz.   In addition to effect of the filter, flux noise also reduces $T_{2,\mathrm{CPMG}}$ at lower frequencies. (b) Single-qubit gate error as measured with randomized benchmarking compared to the coherence limited predicted from coherence times in (a). Negative effects of the filter on qubit operations should appear as an excess of errors with respect to the coherence limit.  Over the range of the LRU, from 4.6 to 4.8~GHz, the error can be well predicted by the coherence limit calculated using \cref{eq:epc}, meaning that we do not detect an excess of gate errors.   
 } 
	\label{fig:lifetime_epc}
\end{figure}

To study the effect of the filter on a qubit as a function of the $eg$-qubit transition frequency we made the qubit flux-tunable via a symmetric dc-SQUID. The results presented in this section are obtained with device A.  \cref{fig:lifetime_epc}(a) shows the results of a measurement of  $T_{1,e}$, $T_{2,\mathrm{CPMG}}$, and $T_{1,f}$, as we sweep the qubit frequency over \SI{500}{MHz} in \SI{10}{MHz} steps from the maximum frequency of $\omega_{ge,\mathrm{max}}/2\pi$ = 4.970~GHz to a minimum frequency of 4.470~GHz.  At each step we recalibrate the gates, and then measure the coherence times and perform randomized benchmarking.  Reported values are averaged over 19 sweeps, taken over the course of 5 days.  As discussed in the previous section, the passband of the filter begins at 4.5~GHz, therefore the maximum qubit frequency is deep in the stopband of the filter.  Furthermore, because the anharmonicity of this transmon is $\alpha/2\pi$ = -325~MHz, the maximum $ef$-transition frequency, $\omega_{ef}/2\pi$ = 4.645~GHz, is also in the stopband of the filter.  Therefore at the sweet spot of the SQUID both the $e$ state and the $f$ state are long lived, with $T_{1,e}$ = 119~$\mu$s and $T_{1,f}$ = 34~$\mu$s.

As we apply flux, we bring both $\omega_{ge}$ and $\omega_{ef}$ towards the passband of the filter.  As the qubit frequency decreases the qubit remains long-lived with $T_{1,e} \approx 100~\mu$s, until  $\omega_{ge}/2\pi \approx$ 4.6~GHz, where the qubit transitions enter the passband of the filter.  We stop measuring at $\omega_{ge}/2\pi$ = 4.470~GHz, because the $T_{1,e}$ is reduced to 1~$\mu$s, at which point qubit calibrations and measurement become difficult. Moreover, tuning the qubit frequency via flux makes the qubit  succeptible to flux noise, which slowly reduces $T_{2,\mathrm{CPMG}}$ as the qubit frequency is tuned away from the sweet spot.  For a fixed-frequency transmon we do not expect the filter to have an effect on the dephasing rate, see Section \ref{sec:dephasing}. 
However, as $\omega_{ef}$ decreases, $T_{1,f}$ is rapidly suppressed. A change in frequency of only 200~MHz causes $T_{1,f}$ to decrease by more than three orders of magnitude, from a maximum of 34~$\mu$s to a minimum $T_{1,f}$ = 33~ns at $\omega_{ge}/2\pi$ = 4.745~GHz ($\omega_{ef}/2\pi$ = 4.417~GHz).  The extremely sharp roll-off of the filter can be demonstrated by noting that at this operating point where the $ef$-transition has been maximally suppressed by the filter, the lifetime of the $e$ state is unaffected, with $T_{1,e}$ = 106~$\mu$s.  
As the frequency continues to decrease the lifetime of the $f$ state remains suppressed below 100~ns, with visible ripple due to the density of states of the filter. This filter can act as an LRU over a 200~MHz bandwidth, for qubit frequencies from 4.6 to 4.8~GHz.  For other filters, this operational bandwidth will vary based on background $T_{1,e}$, desired $T_{1,f}$, anharmonicity of the qubit, and the order of the filter. Moreover, the operational bandwidth of this filter is large enough that non-flux-tunable transmons can be used, as the bandwidth is larger than the spread of qubit frequencies due to junction variability \cite{zhang2022high}. 

Next we perform single-qubit randomized benchmarking (RB), as shown in Fig. \ref{fig:lifetime_epc}(b), to further characterize the effect of the filter on single-qubit gate operations. One potential source of gate errors during single qubit gates in weakly anharmonic qubits like transmons due to the filter is that the $f$ state is transiently occupied during the gate, because of the finite pulse bandwidth. DRAG pulses with a duration of 4 standard deviations are used to reduce the population in the $f$ state at the end of the pulse, but DRAG does not eliminate the population of the $f$ state during the gate \cite{chow2010optimized, chen2016measuring}.  
To test the potentially detrimental impact of relaxation on the transient $f$ state population during the gate, we purposefully try and maximize the $f$ state population by minimizing the duration of the pulse by choosing a total gate time $t_g$ = 14.2~ns for both the $X$ and $Y$ gates. The $Z$-gate is performed with a virtual frame change \cite{mckay2017efficient}. We need to compare the measured RB values to the best RB we could expect to do based on the coherence of the qubit. We use the qubit coherences, $T_{1,e}$ and $T_{2,\mathrm{CPMG}}$, shown in \cref{fig:lifetime_epc}(a) to compute the coherence limit of the single-qubit gate fidelity using \cite{wei2024characterizing}
\begin{align}\label{eq:epc}
\mathcal{E} = \frac{1}{2}\left( 1 - \frac{2}{3}e^{-t_g/T_{2,\mathrm{CPMG}}} - \frac{1}{3}e^{-t_g/T_{1,e}}\right).
\end{align}
Note that this formula does not depend on $T_{1,f}$, and is not modified to account for any potential effects of the filter. Therefore detrimental effects from the filter on $\mathcal{E}$ will show up as an offset between the experimentally measured gate fidelities and the coherence limit.

We see in \cref{fig:lifetime_epc}(b) that in general the error rate increases as the qubit frequency decreases, at first slowly due to the flux-noise reduced $T_{2,\mathrm{CPMG}}$ and then rapidly due the reduced $T_{1,e}$ as the qubit frequency enters the filter.  At the highest qubit frequencies, above 4.8~GHz, the experimental RB fails to achieve the coherence limit.  However, in this regime the filter is not heavily suppressing $T_{1,f}$, so this is not due to the filter. There are many potential reasons for this, such as the very short gate time which could mean that there is residual coherent error, and interactions with strongly coupled two-level systems are also not accounted for by \cref{eq:epc}. However, in the range of qubit frequencies over which we want to use the filter as an LRU, from 4.6~GHz to 4.8~GHz, the experimentally measured single qubit gate fidelities are well predicted by the coherence limit.  Therefore, despite making an aggressive filter to suppress $T_{1,f}$ to 33~ns, and making the gates only 14~ns long to populate the $f$ state as much as possible during a gate, while maintaining a qubit lifetime of 106~$\mu$s, we do not see a suppression of the single qubit error rate to the resolution of the experiment, about \num{1e-4}.  Nevertheless, because the $f$ state is transiently populated, the filter should have an effect on the gate fidelities. Therefore in Section \ref{sec:filter_theory} we theoretically explore analyze a qubit connected to the filter to understand the fundamental limits that the filter imposes on the gate fidelities. 


\subsection{Filter as an LRU }

\begin{figure*}[t!]
	\includegraphics[width=\textwidth]{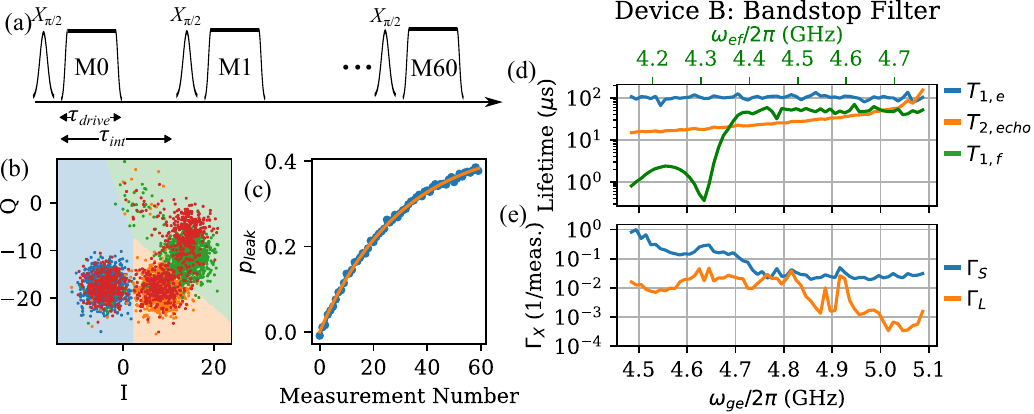}
	\caption{Demonstration of filter as an LRU. (a) Experiment to measure leakage and seepage rates per measurement using a mock-up single-qubit QEC-cycle.  We repeat the sequence $\left( X_{\pi/2} - Meas. \right)$ back-to-back 60 times.  The total readout integration time $\tau_{int}$ = 300~ns, although the resonator is only driven for $\tau_{drive}$ = 142~ns. (b) The outcomes of interleaved calibrations of the $g$ (blue), $e$ (orange), and $f$ (green) states.  The blue, orange, and green backgrounds, show how subsequent measurements get binned into $g$, $e$, or $f$ according to the nearest calibration state.   The red points show the outcomes of the 60th measurement, showing occupation both of the states in the qubit subspace, and occupation of several leakage states. Note that most of the leakage states, even the leakage states above $f$, get binned into the $f$ state during this process. Although 5000 points were taken during each experiment, only 1000 points are shown to reduce clutter.  (c) The blue points show the probability of the $m$th measurement being binned into a leakage state, $p_{\mathrm{leak}}$.  The leakage ($\Gamma_L$) and seepage ($\Gamma_S$) per measurement are can be extracted from the fit (orange) to Eq. \ref{eq:pleak}. (d) $T_{1,e}$, $T_{2,echo}$, and $T_{1,f}$ as a function of $\omega_{ge}$ as measured according to the procedure used in Fig. \ref{fig:lifetime_epc}.  (e) Median leakage and seepage rates per measurement as extracted from the fits to Eq. \ref{eq:pleak}. Notice how the seepage rate increases around $\omega_{ge}/2\pi$ = 4.7~GHz as $\omega_{ef}$ enters the passband of the filter, reducing the lifetime of the last leakage state. 
    } 
	\label{fig:lru}
\end{figure*}

So far we have focused on demonstrating that the  filter does not negatively affect the performance of single qubit gates. We next demonstrate the usefulness of the filter as an LRU. To do this, we use a circuit that simulates the sort of leakage that would occur in a QEC experiment. QEC relies on repeated measurements of check qubits, and readout is one of the leakiest qubit operations \cite{sank2016measurement, khezri2022measurement, shillito2022dynamics, sundaresan2023demonstrating}. Therefore we perform a single-qubit mock-up of a QEC cycle, consisting of the sequence $\left( X_{\pi/2} - \mathrm{Meas} \right)$, repeated back-to-back 60 times, see \cref{fig:lru}(a).  Because the measurements are repeated back-to-back, leakage accumulates over the course of a single run of the sequence, making even small amounts of leakage easy to measure.
This sequence is then repeated 5000 times, with 2~ms between runs, to ensure that leakage from one sequence has relaxed before the start of the next sequence.
From this experiment we can extract the leakage and seepage rate per measurement, as previously demonstrated in Ref. \cite{sundaresan2019interacting}. 

In the device shown in Fig. \ref{fig:lifetime_epc}(a) the $T_{1,f}$ is below 100 ns for much of the range of flux biases.  This extremely short lifetime makes it very challenging to measure the population in the $f$ state.  Therefore to characterize the performance of the LRU, we switched to device B,  which is a bandstop rather than a bandpass  implementation of the filter.  The qubit has an upper sweet spot $\omega_{ge}/2\pi$ = 5.087~GHz where $T_{1,e}$ = 103~$\mu$s and $T_{1_f}$ = 51.7~$\mu$s.  Below $\omega_{ge}/2\pi$ = 4.7~GHz, $\omega_{ef}$ enters the passband of the filter.  The minimum $T_{1,f}$ = 361~ns occurs at $\omega_{ge}/2\pi$ = 4.635~GHz. The reduction of $T_{1,f}$ in device B is not as extreme as in device A, because this qubit is not as strongly coupled to the filter as the in the previous device. The aggressive suppression of $T_{1,f}$ in device A is engineered to study potential negative effects of the filter on single-qubit gates. In contrast, the less aggressive suppression of $T_{1,f}$ in device B is appropriate to act as an LRU on a quantum computer with a QEC cycle time of $\sim$ 1~$\mu$s. 

To make the leakage easier to study, the readout is deliberately tuned up to be likely to induce leakage, by choosing a fast measurement time with a large drive amplitude.The dispersive shift of this qubit is $\chi/2\pi$ = 1.0~MHz at the flux sweet spot, and is connected to a readout resonator of frequency $\omega_r/2\pi$ = 7.309~GHz with a linewidth $\kappa/2\pi$ = 5.5~MHz. The readout pulse has a duration $\tau_{\mathrm{drive}}$ = 142~ns, and the measurement integration time is $\tau_{\mathrm{int}}$ = 300~ns. At the measurement amplitude used in this experiment we are unable to reliably calibrate the number of photons in the resonator. The readout is assisted by both a traveling-wave parametric amplifier on the mixing chamber stage and by the use of a matched filter trained on measurements of the $g$ and $e$ states \cite{ryan2015tomography}.

Interleaved with the sequence shown in \cref{fig:lru}(a), we also perform a calibration consisting of measuring the qubit when prepared in each of the $g$, $e$, and $f$ states.
Figure \ref{fig:lru}(b) shows the IQ blobs for the calibration states.  Each of the 60 measurements in the experiment is then assigned to $g$, $e$, or $f$ according to which calibration has the nearest median point as shown by the shaded regions in Fig. \ref{fig:lru}(b).  The red points in Fig. \ref{fig:lru}(b) show the results of the 60th measurement, showing that the qubit population has spread beyond the $g$ and $e$ states to occupy multiple leakage states. We do not distinguish different leakage states, with all of them being assigned to the $f$ state in our analysis. This data is taken at the upper sweet spot of the qubit, where the $f$ state is long lived because the $ef$-transitions is in the stopband of the filter.  Therefore, at this flux bias we can easily see a clear separation between the $e$ and $f$ states. However, as $T_{1,f}$ is reduced, relaxation from $f$ to $e$ reduces the distinguishability of the $e$ and $f$ states.  To correct for relaxation during readout and overlap of the IQ blobs, we adapt the readout error mitigation technique from Ref. \cite{bravyi2021mitigating} for multiqubit readout and apply it to single-qubit three-level readout.  
To do so, we create a measurement-error mitigation matrix that consists of the probability of each measurement outcome for each of the initial states $g$, $e$, and $f$, assuming perfect initialization in the calibration measurements.  We then apply the inverse of this matrix to the ensemble of outcomes for each iteration of the measurement, to undo measurement errors. 
Note that at the upper sweet spot of device B, the filter does not act as an LRU because $\omega_{ef}$ and $\omega_{fh}$ are not in the passband of the filter.  Therefore  we observe what appears to be a large population in the $h$-state in \cref{fig:lru}(b), because the higher leakage transitions are in the passband of the filter. 

Figure \ref{fig:lru}(c) shows $p_{\mathrm{leak}}$, the probability of each of the measurement being in a leakage state, for each of the 60 measurements in the sequence in Fig. \ref{fig:lru}(a). This probability grows rapidly as population builds up in the leakage states. As population in the leakage states accumulates, eventually the population approaches a steady state where the amount of leakage and seepage is equal.  We fit to the function 
\begin{align}
p_{\mathrm{leak}} = \frac{\Gamma_L}{\Gamma_L+\Gamma_S} \left( 1- \exp^{-\left(\Gamma_L +\Gamma_S \right)m} \right) 
 \label{eq:pleak}
\end{align}
for measurement number $m$, where $\Gamma_L$ is the leakage rate per measurement and $\Gamma_S$ is the seepage rate per measurement. The curve in \cref{fig:lru}(c) is a fit yielding the rates $\Gamma_L$ = 1/67.9~measurements and $\Gamma_S$ = 1/54.9~measurements.  The purpose of the $X_{\pi/2}$ in the circuit is to randomize the state of the qubit between measurements.  It has no effect on the transmon if one of the leakage states is occupied. This simplifies the analysis because the leakage rates from $g$ and $e$  can be very different \cite{khezri2022measurement, dumas2024unified}.  In our procedure, we measure a single leakage rate per measurement averaged over the rates from $g$ and $e$.

Figure \ref{fig:lru}(e) shows the median $\Gamma_L$ and $\Gamma_S$ as a function of qubit frequency averaged over 19 runs.  There is structure in both $\Gamma_L$ and $\Gamma_S$ as a function of frequency.  The dependence of $\Gamma_L$ on qubit frequency has been a subject of recent study both experimentally and theoretically \cite{khezri2022measurement, dumas2024unified}.  Here we focus on the structure in $\Gamma_S$. We see that above $\omega_{ge}/2\pi$ = 4.8~GHz, $\Gamma_S$ is fairly stable versus frequency. Below $\omega_{ge}/2\pi$ = 4.7~GHz, $\Gamma_S$ increases as $\omega_{ef}$ enters the passband of the filter, as can be seen in $T_{1,f}$ in Fig. \ref{fig:lru}(d).  We see a local maximum in $\Gamma_S$ at $\omega_{ef}/2\pi$ = 4.64~GHz, corresponding to the minimum in $T_{1,f}$.  The correspondence between $\Gamma_S$ and $T_{1,f}$ shows that the filter is responsible for this increase in seepage back into the qubit subspace. Because measurement-induced leakage is often to states much higher than the $f$ state, requiring a multi-stage relaxation process to return to the qubit subspace, it is likely that the filter is having an effect on the relaxation of the higher states prior to $\omega_{ef}$ entering the passband of the filter.  However, because the $f$ state is likely to have the longest lifetime of the leakage states, $\omega_{ef}$ entering the passband of the filter has the largest effect on $\Gamma_S$.

\section{Filter Theory}\label{sec:filter_theory}
In this section we provide a theoretical description of the device. First, we briefly review how we obtain a quantum-mechanical description of the bandpass filter under consideration. We then explain how we perform simulations of the coupled qubit-filter system. Finally, we discuss how one expects the filter to impact the performance of logical operations on the qubit, and show numerically that high-fidelity gates are possible in the presence of this engineered dissipation. 

\subsection{Quantum description of a bandpass filter}
We first summarize how to obtain a quantum description of a bandstop filter coupled to two output ports. In the next subsection, we replace the first port by the qubit we wish to protect from leakage. Although the physical filter is implemented as a coupled-line filter as shown in Fig. \ref{fig:filter_phys}(b), we quantize the much simpler but functionally-equivalent lumped-element implementation of the filter as shown in Fig. \ref{fig:filter_phys}(a).

As was previously mentioned, there is a standard design flow to build such a filter \cite{pozar2011microwave}. Given a set of target filter parameters --- which for a Chebyshev filter of order $L$ are the central frequency $\omega_0$, the bandwidth $\Delta \omega$ and the ripple factor $\eta$ --- one obtains a set of normalized element values $\{g_n\}$ which correspond to capacitances, inductances, and impedances of a lumped-element circuit.  Quantizing this model through a standard procedure \cite{naaman2022PRXSynthesis}, one obtains a 1D chain of nearest-neighbour coupled linear bosonic modes. The length $L$ of the chain matches the order of the filter. The corresponding second-quantized Hamiltonian within the rotating-wave approximation (RWA) is 
\begin{align}\label{eq:H_f}
\hat{H}_f
=
\omega_0
\sum_{n=1}^L
\hf_n^\dagger \hf_n
+
\sum_{n=1}^{L-1}
J_{n,n+1}
\left(
\hf^\dagger_{n+1} \hf_n
+
\rm{h.c.}
\right),
\end{align}
where $\hf_n$ is a the mode annihilation operator of site $n$. The coupling constants $J_n$ are directly related to the design coefficients \cite{naaman2022PRXSynthesis}, which for the filter under consideration takes the form
\begin{align}\label{eq:J_n}
J_{n, n+1}  &= \frac{\Delta \omega}{4}
\frac{|\sin(n \frac{\pi}{L} + i \beta )|}
{
\sqrt{
\sin \frac{(2n-1)\pi}{2 L}
\sin \frac{(2n+1)\pi}{2 L}
}
},
\end{align}
where $\beta = \textnormal{arcsinh}(\eta^{-1})/L$. 

 Momentarily ignoring the qubit, the first and last sites are coupled to input-output waveguides which defines port 1 and 2 of the device, respectively. This induces Markovian damping on those two sites at a rate $\kappa_f$, whose form is also determined by $\{g_n\}$ and here is given by
\begin{align}\label{eq:kappa_f}
\kappa_f
=
\frac{\Delta \omega}{2}
\frac{\sinh \beta}{\sin \frac{\pi}{2 L}}.
\end{align}
Focusing exclusively on the modes $\hf_n$ of the filter, their dynamics is described by a Lindblad master equation with Hamiltonian $\hat{H}_f$ and dissipators $\kappa \mathcal{D}[\hf_1] \hrho + \kappa\mathcal{D}[\hf_L] \hrho$, with $\mathcal{D}[\hat{X}] \cdot = \hat{X}\cdot \hat{X}^\dagger - \{\hat{X}^\dagger \hat{X}, \cdot\}/2$. If our focus was, however, on how signals entering port 2 are transmitted to port 1, one would use quantum input-output theory to compute the frequency-dependent scattering matrix element $S_{12}[\omega]$ \cite{clerk2010introduction}. The upshot is that with our choice of Hamiltonian \cref{eq:H_f} and parameters \cref{eq:J_n}-(\ref{eq:kappa_f}), the scattering matrix obtained from the fully quantum calculation exactly matches the classical scattering matrix 
\begin{align}
|S_{12}[\omega]|^2
=
\frac{1}
{1+
\eta^2
T_{L}^2
(
\frac{2 (\omega-\omega_0)}{\Delta \omega}
)
},
\end{align}
where $T_{L}$ is the Chebyshev polynomial of the first kind of order $L$. Unless otherwise stated explicitly we fix $L = 7$, although the following discussion easily generalizes to a filter of arbitrary order. 

\subsection{Coupling the qubit to the filter}
With a quantum description of the filter in hand, we now bring back the coupling to the transmon qubit. We do so by replacing the Markovian input-output waveguide on the first site by the qubit, and removing the term $\kappa \mathcal{D}[\hf_1]$ from the master equation. This term is replaced by a capacitive coupling between the transmon and a resonator which enters the Hamiltonian. In what follows, we will drop the qubit's intrinsic decay and dephasing rate for the sake of compactness; we keep them in all numerical simulations throughout this work. With that caveat, the master equation describing the full qubit-filter density matrix then takes the form 
    \begin{align}\nonumber
    	\partial_t \hrho 
           =
    	&-i
    	[
    	\hH_t + 
            \hH_f 
            +
            \epsilon_t(t)\hn_t
    	-i g\hat{n}_t\left(\hf_1-\hf_1^\dagger \right)
    	,
    	\hrho
    	]
     \\ \label{eq:Full_EOM}
    &+
    \kappa_f
    \mathcal{D}[\hf_7]
    \hrho.
    \end{align}
Here, $\hat{H}_t = 4 E_C \hn_t^2  -E_J \cos \hphi_t$ is the usual transmon Hamiltonian, with $E_C$ the charging energy, $E_J$  the flux-tuneable Josephson energy, $\hn_t$, and $\hphi_t$ the standard charge and phase operators of the transmon satisfying the usual canonical commutation relations $[\hphi_t, \hn_t] = i$. As is standard, we have ignored any static gate charge on the transmon \cite{blais2021circuit}. The function $\epsilon_t(t)$ is the pulse shape of the  charge drive that we will use to perform gates. The only source of decay comes from coupling the last site $n=7$ to a lossy element, which is a consequence of assuming the filter elements are purely reactive. Finally, $g$ is the transmon-filter coupling, which we note is the only free parameter in the model. We use this free parameter to match the experimentally-measured decay rate of the $f$ state, and for our parameters is $g/2\pi = \SI{20}{MHz}$. 

The master equation \cref{eq:Full_EOM} constitutes the full quantum description of the coupled qubit-filter system. Beyond brute-force numerical simulation of \cref{eq:Full_EOM}, which at first glance seem rather challenging given the numerous bosonic modes, there are several other ways to analyze the dynamics. First, one could think of the filter as a highly-structured environment coupled to the qubit, i.e.~a non-Markovian bath. Following this line of reasoning, one could integrate out the linear filter modes exactly and obtain a non-Markovian equation of motion for the qubit only. However, non-Markovian master equations are more difficult to work with numerically \cite{Rosenbach_2016}, and the theory of these master equations is not as nearly well-developed compared to their Markovian counterparts \cite{breurerpetruccione}.

Another option to capture the effects of the filter on the qubit is perturbation theory. By assuming the coupling $g$ is weak enough, one could make the approximation that the density of states (DOS) at any transition  frequency is locally flat, and that this is sufficient to capture its dissipative effects \cite{breurerpetruccione}. One would be left with a Markovian master equation with dissipators describing different transitions in the transmon at different rates determined by Fermi's golden rule. However, the ideal operating range of the filter is precisely when the $f$ state is near the edge of the filter's passband,  where the DOS is decidedly not flat, see Fig.~\ref{fig:filter_phys} (d). Further,  the Markovian approximation is uncontrolled. As a result, looking ahead towards implementing this architecture in a quantum computer where one aims to achieve gate fidelities on the order of $99.99\%$, a numerically-obtained averaged gate fidelity to this level of precision using this Markovian approximation cannot be trusted. Without a reason \textit{a priori} to believe in perturbation theory, to reach this level of precision we are forced to treat the complete dynamics governed by the master equation \cref{eq:Full_EOM}.

\begin{figure}
    \centering
    \includegraphics[width=0.47
\textwidth]{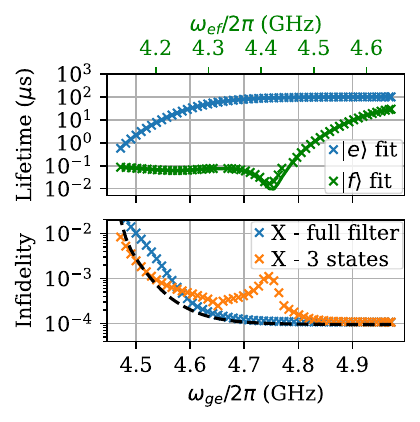}
    \caption{a) Decay rate of the qubit's $e$ and $f$ state, obtained either via a fit to the full master equation \cref{eq:Full_EOM} (coloured crosses) or Fermi's Golden Rule \cref{eq:FGR} (full line). Only slight deviations between the two appear when the $\omega_{ef}$ enters the filter, where we except the perturbative expression to fail in any case. (b) Averaged infidelity of a DRAG-optimized $\SI{14.2}{ns}$ $X$ gate (solid line) using the full filter (blue) and keeping only the $g,e$ and $f$ state of the qubit (orange) with corresponding decay rates from (a). The discrepency between the two is large when the DOS at the $ef$ transition enters the filter and the Markovian approximation is expected to fail. The identity gate (dashed black line), also referred to as the coherence limit, for the same duration is also plotted.  We see that over a large range of frequencies $\sim \SI{100}{MHz}$ we have a long-lived $e$ state lifetime $\sim \SI{100}{\mu s}$, short-lived $f$ state lifetime $\sim \SI{100}{ns}$ and low single qubit gate infidelity $\sim 10^{-4}$. Further, all relevant qualitative features in these two theory plots match their experimental counterparts in \cref{fig:lifetime_epc}. As one further decreases the qubit frequency it enters the filter and starts to hybridize with the filter modes. This explains the largening discrepency between the two gate fidelities as we lower the qubit frequency. These numerical simulations were run with a maximum number of excitation $N_{\rm exct}$ = 2. Increasing this dimension to $N_{\rm exct} = 3$ and $N_{\rm exct} = 4$ (not shown) does not change the result, the averaged infidelity between any of the three simulations differing by $\sim 10^{-7}$.}
    \label{fig:T1_numeric}
\end{figure}

\subsection{Partitioning the Hilbert space}

While it is well known that measurement and other strong microwave drives can populate very high states of the transmon \cite{sank2016measurement, khezri2022measurement, shillito2022dynamics}, to understand the effect of the filter on the gates, here we focus on the states $g$, $e$, and $f$.  This is because only the $f$ state has a significant transient population during the gate, and we want to study the effect of the filter on single qubit gates. Since excitations can be exchanged between the transmon and filter via the coupling $g$, we must also keep the same number of levels in all other filter states, that is up to $2$ photons in each mode. The Hilbert space dimension of the coupled system for the $7$-th order Chebyshev filter is thus at minimum $3 \times 3 ^7 \sim 10^4$. Although it is possible to numerically time-evolve master equations with Hilbert spaces of this size, performing any sort of parameter exploration or optimizing over gates is numerically too demanding. In addition, eventually including more leakage levels in the transmon  would only makes this problem worse. It is thus clear that a sheer brute-force approach to solving \cref{eq:Full_EOM} is not practical. 

We can, however, use the structure of the master equation to judiciously eliminate certain states which will be irrelevant to the dynamics. Indeed, within the RWA the Hamiltonian $\hat{H}_t + \hat{H}_f -i g \hat{n}_t (\hf_1-\hf_1^\dagger)$ preserves the total excitation number. Further, although the dissipation is strong, it can only remove excitations from the system. Thus if we start in e.g. the $f$ state of the transmon with no photons in the filter, $|f, 0^{\otimes 7}\rangle $, we can safely ignore states with one photon in each filter mode such as $|e, 1^{\otimes 7} \rangle$. This state has 8 excitations, and can be safely dropped from the dynamics. Defining $N_{\rm exct}$ as the maximum total excitation number we will keep in our time evolution, a simple combinatorial calculation reveals that the dimension of this new partitioned Hilbert space is $\binom{1+7+N_{\rm exct}}{N_{\rm exct}}$. For $N_{\rm exct} = 2$, this corresponds to a Hilbert space dimension of $45$. We can then easily perform these simulations in this reduced Hilbert space and confirm the validity of our results by verifying that they are insensitive to an increase in $N_{\rm exct}$.

\subsection{Effects of the filter on qubit decay rates}
We are now in a position to explore the effects of the filter on the qubit. By design, at the operating point the transmon is weakly-coupled to the filter and the $ge$ transition is in the stopband. The qubit subspace is then safely in the dispersive regime. This leads to a weak dressing of the qubit eigenstates which we denote $|\overline{g, 0^{\otimes 7}} \rangle$ and $|\overline{e, 0^{\otimes 7}} \rangle$. These numerically-obtained states serve as our computational basis. The coupling also leads to a small dressing of the qubit's frequency, in complete analogy to the standard coupling between a transmon and a resonator \cite{blais2021circuit}. We denote $\bar{\omega}_{ge}$ as the bare qubit frequency and $\omega_{ge}$ as the dressed qubit frequency, which note is the opposite notation for states.

To estimate the decay rate of our qubit's excited state and the leakage states, we can use Fermi's Golden Rule \cite{Sakurai_Napolitano_2020}. Under the usual RWA,  
the decay rate of the qubit $\Gamma_{e \to g}$ and first leakage state $\Gamma_{f \to e}$ induced by the filter take the form
\begin{align}\label{eq:FGR}
\begin{split}
\Gamma_{e\to g}
&=
2\pi
|g_{eg}|^2
\rho_{f}
(
\bar{\omega}_{ge}
),
\\
\Gamma_{f\to e}
&=
2\pi
|g_{fe}|^2
\rho_{f}
(
\bar{\omega}_{ge}+\alpha
),
\end{split}
\end{align}
where $g_{i_t j_t} \equiv g \langle i_t | \hn_t |j_t\rangle$, $\rho_{f}(\omega)$ is the local density of states (LDOS) of the first site of the filter, and recall that $\alpha < 0 $ is the transmon anharmonicity. The LDOS is related to the linear response properties of the filter and, in particular, can be determined by its frequency-dependent retarded Green's function
\begin{align}
G^R_f[n,m;\omega]
&=
-i \int dt
e^{i\omega t}
\theta(t)
\langle [\hf_{n}(t), \hf_{m}^\dagger(0)] \rangle,
\\
\rho_f(\omega)
&=
-\frac{1}{\pi}
\mathrm{Im}
\:
G^R_f[1,1;\omega].\label{eq:DOS}
\end{align}
Here the filter creation and annihilation operators are in the Heisenberg picture with $g = 0 $, $\theta(t)$ is the Heaviside step function and the average is with respect to the steady-state of the filter. The function $G^{R}_f(n,m;\omega)$ characterizes the frequency response of site $n$ under a force acting on site $m$. It is worth emphasizing that for the linear filter considered here the density of states can be computed directly from coupled-mode theory and is directly related to the filters impedance, see Refs.~\cite{naaman2022PRXSynthesis, yan2023broadbandpurcell} for details.

In \cref{fig:T1_numeric} we plot the predictions of Eq.~(\ref{eq:FGR}) over a range of qubit frequencies and compare with the numerically-extracted decay rates obtained by solving \cref{eq:Full_EOM}. We see an excellent agreement between the two methods over essentially the whole range of frequencies. Placing the qubit's transition frequency in the stopband is thus sufficient to protect it from unwanted additional decay. We stress that being able to predict the correct decay rates via perturbation theory does not imply that we can obtain the infidelities using this method, i.e. with a Markovian description of the qubit.

\subsection{Effects of filter on qubit dephasing rate}\label{sec:dephasing}
Another relevant source of error that can be enhanced by the filter is dephasing. To understand why, consider the standard dispersive readout of a transmon coupled to a readout resonator with mode annihilation operator $\hat{a}$ leading to dispersive interaction $\chi \hat{\sigma}_z \hat{a}^\dagger \hat{a}$. It is well-known that shot-noise fluctuations of the resonator photon number under this dispersive coupling leads to enhanced dephasing \cite{blais2004cavity}.
 
Naively, this effect is even worse here, where we purposefully make the $ef$ transition resonant with the filter: the analogue to the $\chi$ shift here should thus be large, given that its denominator precisely involves the detuning between these two states \cite{koch2007transmon}. It seems as though the presence of the filter necessarily involves an increase in the pure dephasing rate of the qubit. 

Crucially, however, it is fluctuations in $\ha^\dagger \ha$ which causes additional dephasing of the qubit. Consequently, any enhanced dephasing on the qubit due to the filter should vanish if the latter is in vacuum. By adding weak incoherent pumping of the form $\kappa_f \bar{n}_{\rm th}\left(\mathcal{D}[\hf_7] + \mathcal{D}[\hf_7^\dagger] \right) \hrho$ to \cref{eq:Full_EOM} we verify numerically both of these effects, see Appendix~\ref{app:n_th}. Namely, we show that the pure dephasing rate of the qubit is largely unaffected by the presence of the filter at zero temperature and, further, that the pure dephasing rate is indeed made worse when the $ef$ transition enters the bandpass of the filter. This is confirmed by the experimental data presented in Appendix~\ref{app:n_th}, where a decrease in the $T_2$ time is observed when the filter's temperature is purposefully increased.

\subsection{Effects of the filter on single-qubit gates}
Having characterized how the filter affects the qubit's decay and dephasing rates, we can now address how single-qubit gate fidelities are changed in the presence of the engineered dissipation. Before addressing this question directly, however, we first examine how the filter affects the transient $f$ state population during a gate. As we have emphasized, since the qubit subspace is largely unaffected by the presence of the filter, this should be the leading-order effect if the filter were to have any deleterious effects on single qubit operations.

\begin{figure}
    \centering
    \includegraphics[width=0.47
\textwidth]{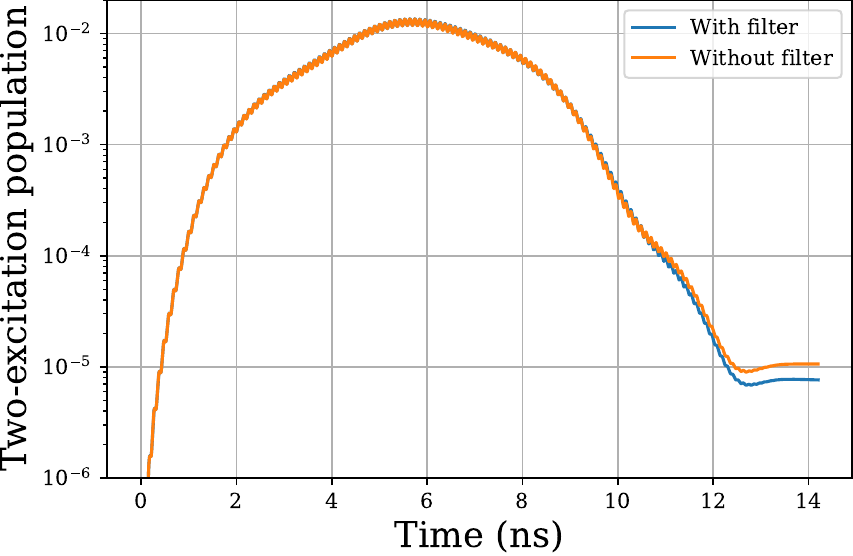}
    \caption{Numerical simulation} comparing the transient two-excitation population during a 14.2~ns long DRAG-optimized $X$-gate with and without the filter. The qubit frequency is such that the $f$ state lifetime obtained from the numeric fit $T_{1,f} \approx 9. 89$ ns is minimal. We see essentially no difference between the two plots during the bulk of the gate, and only a very minor difference between the two at the end. 
    \label{fig:f_state_pop}
\end{figure}

In Fig.~\ref{fig:f_state_pop} we plot the population of the two-excitation states (which includes the $f$ state with no photons in the filter) during a DRAG-optimized $X$-gate, obtained via a numerical integration of Eq.~(\ref{eq:Full_EOM}). Here, the DRAG parameters are chosen to optimize the average fidelity of the $X$-gate. We see that this population throughout the gate is largely independent of whether or not we have an engineered dissipative environment. It is worth emphasizing that the objective of the filter is not to remove any of this transient population during the gate. Rather, the purpose of this figure is to show that the impact of the filter on the $f$ state population should be minimal. Consequently, we expect the impact on single-qubit gates should also be negligible.

This expectation is borne out in Fig.~\ref{fig:T1_numeric}(b) where we plot the average infidelity of the $X$ gate and identity gate --- also known as the coherence limit --- as a function of qubit frequency. For each qubit frequency we choose the DRAG parameters to optimize this averaged infidelity. At the operational point of the filter $\omega_{ge}/2\pi \approx 4.7 \sim 4.8$GHz, where we have both a large decay rate of the $f$ state and large $T_{1,e}$ we obtain an averaged infidelity $\sim 10^{-4}$ which is close to the coherence limit. This demonstrates that the filter does not have any major detrimental effects to single-qubit operations. This is, however, not the case if we were to simply describe the effects of the filter on the qubit within a Markovian approximation, see the orange line in Fig.~\ref{fig:T1_numeric}. This is not surprising, given that the DOS near the range of frequencies is not flat, see Fig.~\ref{fig:filter_phys} d).

As one lowers the qubit frequency, we see a large difference between the fidelity of the $X$ gate and the coherence limit, unlike what is seen in experiments, see Fig.~\ref{fig:lifetime_epc}. For this range of frequencies, the dressed states we define as our computational subspace become more and more filter-like. In a worst-case situation for instance, the dressed qubit excited state is a superposition $|\overline{e, 0^{\otimes 7}}\rangle \propto |e, 0^{\otimes 7}\rangle + |g,0^{\otimes 6},1\rangle$ while the orthogonal state $\propto |e, 0^{\otimes 7}\rangle - |g,0^{\otimes 6},1\rangle$ would be considered a leakage state.  While our numerical simulations can resolve these two states, the experiment presumably cannot. We thus do not expect an agreement between theory and experiment in this regime.  This is however not in a range of parameters one would want to operate the qubit in.  

\section{Conclusions and Outlook}

We have demonstrated that a high-order filter between a qubit and a source of cold resistance can be used as an LRU.  We have showed that the filter can reduce the lifetime of the first leakage state by three orders of magnitude, while protecting the qubit lifetime. Crucially, high-quality single-qubit operations can be performed in the presence of the filter. Even though the $f$ state is transiently occupied during a fast single-qubit, we do not observe negative effects of decoherence on the $f$ state of the qubit on single qubit gate fidelities, despite the gate time approaching the lifetime of the $f$ state. Our analytical and numerical results support the experimental data, with a strong quantitative agreement between both lifetimes and single-gate fidelities.

Our approach to leakage reduction has several advantages over other experimentally-implemented LRU's. Because it is always on, it does not slow down the computation by adding an extra step for leakage reset. Further, the passive nature of this bath-engineering approach does not require additional control hardware and is compatible with fixed-frequency qubits. Finally, the broadband nature of the filter means that unwanted accumulation of population to higher-leakage state is automatically accounted for. 

Future work is needed before our filter-based approach to building an LRU is suitable in a full-scale quantum computer. First, in this paper we did not address the effect of the filter on two-qubit gates. Some implementations utilize the $f$ state on at least one of the qubits.  In this case, unless the qubit or filter are tunable in frequency, the $ef$-transition would need to be outside the passband of the filter state.  Second, the current implementation of the filter is too large to be scalable.  More compact implementations of filters can be achieved using  spiral resonators \cite{park2024characterization}, nano-mechanical resonators \cite{cleland2019mechanical}, or metamaterial transmission lines \cite{zhang2023superconducting, ferreira2024deterministic, scigliuzzo2022controlling}. An alternate approach would be to move the filter off-chip into a multi-level wiring layer.

\begin{acknowledgments}
We thank Muir Kumph, David McKay, George Stehlik, Oliver Dial, and Ross Shillito for interesting conversations.
T.T. was supported for device characterization by IARPA under LogiQ (W911NF-16-1-0114). 
The views and conclusions contained in this document are those of the authors and should not be interpreted as representing the official policies, either expressed or implied, of the Army Research Office, IARPA, or the U.S. Government. The U.S. Government is authorized to reproduce and distribute reprints for Government purposes notwithstanding any copyright notation herein. Additional support is acknowledged from NSERC, the Canada First Research Excellence Fund, and the Ministère de l'Économie et de l'Innovation du Québec. 
\end{acknowledgments}

\appendix
\section{Thermally-enhanced dephasing}
\label{app:n_th}
In this appendix, we will demonstrate both experimentally and numerically that the qubit experiences excess dephasing when the filter is thermally-populated. This is analogous to the usual shot-noise dephasing of a qubit dispersively-coupled to a resonator, which depends on the $\chi$ shift of the resonator. As we stressed in the main text, a naive perturbative argument suggests that the analogue of the $\chi$ shift should diverge \cite{boissonneault2010improved}, given that a filter mode is resonant with the $ef$ transition. While non-perturbative corrections make the analogue of the $\chi$ shift finite,  crucially the enhanced density of states of the bath that lead to an enhanced $f$ state decay rate also leads to an increase in the pure dephasing rate.

\begin{figure}[t!]
	\includegraphics[width=\columnwidth]{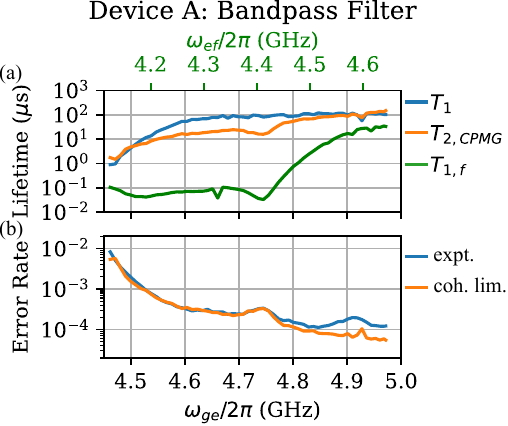}
	\caption{Effect of thermal photons in filter, repeating procedure from Fig. \ref{fig:lifetime_epc}. In this case when the filter is maximally suppressing $T_{1,f}$, there is a dip in $T_{2,\mathrm{CPMG}}$, and a corresponding peak in the single-qubit error rate, indicative of errors from the thermal photons in the filter.  }
	\label{fig:hot}
\end{figure}

Let us begin with the experimental results. To put a thermal photon population in the filter, we exposed the output of the filter to a microwave drive line.  Attenuation on the drive line on the mixing chamber of the fridge provided a 50 $\Omega$ termination for the filter.  However, the attenuation of the drive line was chosen to be insufficient to protect the qubit from the thermal noise from higher temperature stages of the fridge \cite{yan2018distinguishing}. Figure \ref{fig:hot} repeats the experiment from \cref{fig:lifetime_epc} with a hot filter.
While $T_{1,e}$ and $T_{1,f}$ are not affected by the thermal photons, we see additional dephasing in $T_{2,\mathrm{CPMG}}$. There is a dip in $T_{2,\mathrm{CPMG}}$, and peak in the single qubit error rate at $\omega_{ge}/2\pi$ = 4.75~GHz, at the same point where there is a dip in $T_{1,f}$.  This suggests an enhanced dephasing from thermal photons in the filter when the $ef$-transition is near resonant with a mode of the filter. Because this enhanced dephasing potentially affects the operation of the qubit, in this section we seek to understand the source of this dephasing.

\begin{figure}
    \centering
    \includegraphics[width=0.47
\textwidth]{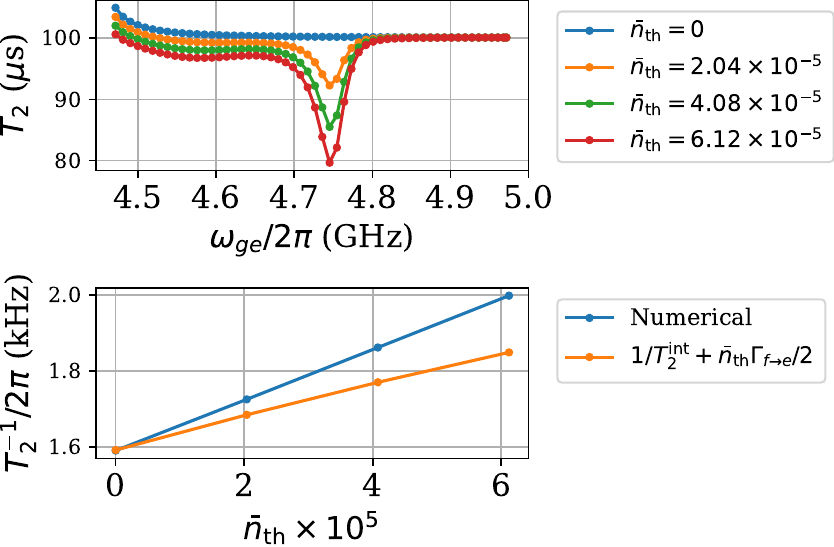}
    \caption{Top : Pure dephasing time $T_{2}$ of the as a function of qubit frequency with and without thermal excitations present in the filter. A thermal occupation of $\bar{n}_{\rm th} = 2.04 \times 10^{-5}$ corresponds to the average occupation of a $\SI{4.5}{GHz}$ bosonic mode coupled to a bath at $\SI{20}{mK}$. We define the pure $T_2$ in the standard way: we start in the superposition state $|\psi(0)\rangle = (|\overline{g, 0^{\otimes 7}} \rangle+|\overline{e, 0^{\otimes 7}} \rangle)/\sqrt{2}$ and fit $T_2$ to the expectation value of the off-diagonal matrix element $|\langle \overline{g, 0^{\otimes 7}} | \hat{\rho}(t)|\overline{e, 0^{\otimes 7}} \rangle|  = \exp(-t\left(1/(2 T_{1,e}) + 1/T_2 \right))/2$. The qubit has an intrinsic $T_2$ of 100$\mu$s. The increase in $T_2$ at lower frequencies is a consequence of the qubit frequency entering in resonance with a filter mode. As the modes hybridize, the dressed qubit becomes more filter-like, which does not suffer from pure dephasing, leading to this slightly-enhanced dephasing rate. Bottom: Pure depahsing rate $T_2^{-1}/2\pi$ as a function of thermal population evaluated at the point where the dephasing time is smallest $\omega_{ge}/2\pi \approx 4.745 \: \rm{GHz}$. Additional dephasing larger than the qubit's intrinsic depahsing time of $T_{2}^{\rm int} = 100 \mu \: \rm{s}$ scales linearly with temperature and is the same order of magnitude as the decay rate of the $f$ state.}
    \label{fig:n_th_T_2}
\end{figure}

To describe the heating of the filter, we assume the thermal photons enter through the last mode, which is coupled to a Markovian bath with an average photon number $\bar{n}_{\rm th}$. The master equation then takes the standard form 
\begin{align}\nonumber
    	&\partial_t \hrho 
           =
    	-i
    	[
    	\hH_t + 
            \hH_f 
    	-i g\hat{n}_t\left(\hf_1-\hf_1^\dagger \right)
    	,
    	\hrho
    	]
     \\ \label{app:EOM_n_th}
    &+
    \kappa_f(\bar{n}_{\rm th}+1)
    \mathcal{D}[\hf_7]
    +
        \kappa_f \bar{n}_{\rm th}
    \mathcal{D}[\hf_7^\dagger]
    \hrho,
    \end{align}
where as in the main text for notational convenience we have suppressed all intrinsic decay and dephasing processes on the transmon, which we keep in all simulations throughout this work. As was stressed in the main text, we expect a finite $\bar{n}_{\rm th}$ to increase the pure depahsing rate $T_{\phi}$ of the qubit when the $ef$ transition enters the filter, a consequence of the enhanced $\chi$ shift at that transition. This is borne out in simulations, where in \cref{fig:n_th_T_2} we compare a numerical fit of the $T_2$ of the dressed qubit $|\overline{g, 0^{\otimes 7}} \rangle$, $|\overline{e, 0^{\otimes 7}} \rangle$ with and without thermal excitations. We see a decrease of the qubit's $T_\phi$ rate around $\SI{4.75}{GHz}$, which is precisely when then decay of the $f$ state is maximal, see \cref{fig:T1_numeric}.

\section{Impact on spectator}

\begin{figure}[b]
	\includegraphics[width=\columnwidth]{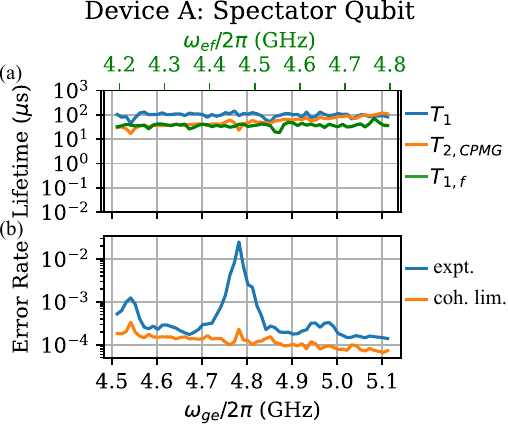}
	\caption{Effect of the filter on a spectator qubit, measured using the same method from Fig. \ref{fig:lifetime_epc}.  (a) Measured $T_{1,e}$, $T_{2,echo}$ and $T_{1,f}$ as a function of qubit frequency, averaged over 19 runs.  (b) Single-qubit error rate as measured with RB, compared to the coherence limit.  The large spike in the middle of the plot is when the spectator qubit collides with the test qubit.      }
	\label{fig:spec}
\end{figure}

So far we have only studied the effect of the filter on a qubit directly coupled to the filter.  To understand the effect of the filter in a realistic quantum computer, we need to understand the effect of the filter on neighboring qubits.  Here we use a second qubit on device A, called the spectator, not directly coupled to a leakage-reset filter. In Fig. \ref{fig:spec}, we repeat the experiment from Fig. \ref{fig:lru} on a spectator qubit coupled with a coupling strength $g/2\pi$ = 1.6~MHz to the test qubit via a short capacitive coupling. To ensure a realistic coupling between the spectator qubit and the filter, we position the test qubit near its optimal working point, where $T_{1,f}$ is minimized. Figure \ref{fig:spec}(a) shows $T_{1,e}$, $T_{2,\mathrm{CPMG}}$, and $T_{1,f}$ on the spectator qubit. If the filter were having an effect on the spectator qubit, we should see a reduction in $T_{1,f}$ in the passband of the filter, below $\omega_{ge}/2\pi$ = 4.75~GHz, however, we see that $T_{1,e}$ and $T_{1,f}$ are flat as a function of frequency.  As expected $T_{2,\mathrm{CPMG}}$ is reduced at lower frequencies because of flux noise.  In Fig. \ref{fig:spec}(b) we show the single qubit error rate as measured with RB.  We see a spike in the error around $\omega_{ge}/2\pi$ = 4.75~GHz, where the spectator and test qubits collide.   

\section{Experimental procedures}

We measure $T_{1,e}$ using the sequence $X_{\pi}-\mathrm{Wait}-\mathrm{Meas}$.  We use the sequence $X_{\pi}-X_{\pi}^{ef}-\mathrm{Wait}-X_{\pi/2}-\mathrm{Meas}$ to measure $T_{1,f}$.  We found that this sequence gave reliable measurements over a broad range of values for $T{1,e}$ and $T{1,f}$.  To measure the dephasing rate, we used a CPMG sequence with $n_{echo}$ = 11, alternating between $X$ and $Y$echo pulses \cite{meiboom1958}.  

\bibliography{bib.bib}

\end{document}